\documentclass[12pt,preprint]{aastex}

\shorttitle{Volatile-Rich Planets}
\shortauthors{Kuchner}

\received{2003 March 9}
\begin{document}

\slugcomment{To appear in ApJ letters October 10, 2003}

\title{Volatile-Rich Earth-Mass Planets in the Habitable Zone}

\author{Marc J. Kuchner\altaffilmark{1}}



\altaffiltext{1}{Michelson Postdoctoral Fellow,
Harvard-Smithsonian Center for Astrophysics,
60 Garden Street, Cambridge, MA 02138, mkuchner@cfa.harvard.edu}

\begin{abstract}

A small planet is not necessarily a terrestrial planet.
Planets that form beyond the snow line
with too little mass to seed
rapid gas accretion ($\lesssim 10$~$M_{\bigoplus}$) should
be rich in volatile ices like H${}_{2}$O and NH${}_3$.
Some of these planets should migrate inward by interacting
with a circumstellar disk or with other planets.
Such objects can retain their volatiles for billions
of years or longer at $\sim 1$ AU as their thick steam atmospheres undergo
slow hydrodynamic escape.  These objects could
appear in future surveys for extrasolar Earth analogs.

\end{abstract}


\section{INTRODUCTION}

Several upcoming space
missions\footnote{See \url{http://planetquest.jpl.nasa.gov}}
aim to detect extrasolar planets
with masses of order 1~$M_{\bigoplus}$.
The Space Interferometry Mission (SIM) should be able to
detect extrasolar planets down to $\sim 3$ $M_{\bigoplus}$
astrometrically by observing the reflex orbital motions of their host stars
\citep{ford03}.
The Kepler and COROT missions are designed to photometrically detect planets
with the same radius as the Earth by observing
the dimming of their host stars during planetary transits.
Eventually, missions like the Terrestrial Planet Finder (TPF)
and Darwin should directly detect extrasolar planets around nearby stars
and take their spectra, searching for planets as faint as
the Earth in either reflected light \citep{kuch03} or thermal emission \citep{wool03}.

Earth-mass planets offer particular biological interest if they
orbit in the habitable zone,
the range of circumstellar radii
where stable liquid water can exist on the surface of a planet
that is in thermal equilibrium with stellar radiation 
\citep{huan59,kast93}.  In the solar system, this
zone covers roughly the region between the orbits of
Venus (0.72~AU) and Mars (1.52~AU).  
Natural places to look for life-bearing extrasolar planets
are the habitable zones of nearby stars.

Discussions of extrasolar planets often
quietly assume that any object with mass
$\sim 1 M_{\bigoplus}$ orbiting in a star's habitable 
zone will be terrestrial, i.e, composed mostly of
silicates and iron-peak elements---like the Earth.
However, we suggest that the habitable zones of nearby stars could
harbor other similar-looking beasts.  Planets composed
substantially of volatiles
can conceivably form with masses up to $\sim 10 M_{\bigoplus}$ and
migrate inward to the habitable zone.
We estimate that Earth-sized volatile-rich planets can retain their volatiles
for billions of years at $\sim 1$~AU,
protected by thick atmospheres that slowly undergo hydrodynamic escape.

\section{FORMATION AND MIGRATION}

The snow line in a protoplanetary disk \citep{haya81}
occurs where the temperature of the nebular gas
drops below the sublimation temperature for water
(170~K in a vacuum). 
In popular planet-formation scenarios \citep[e.g.][]{poll96,wuch00},
planets form much faster beyond the snow line than in the
warmer center of the disk because outside the snow line,
frozen volatiles provide more 
solid building material.  Protoplanets beyond the
snow line more massive than a threshold mass roughly equal to the presumed
core masses of Jupiter and Saturn rapidly
accrete gas, becoming gas giants
\citep[but see also][]{boss02}.
Protoplanets beyond the snow line less massive than this threshold grow slowly,
and when the gaseous disk has dispersed they remain
small balls of ice and rock, like Pluto and the icy moons
of the outer planets.  

Icy planets and moons that form near the snow line should
contain mostly H${}_{2}$O and silicates.  Those that form farther out can
acquire NH${}_3$ and CH${}_4$ as hydrates,
together with CO, N${}_2$, and CO${}_2$.  Beyond the snow line,
the solar nebula does not have time to reach
chemical equilibrium, making it hard to predict the
relative ratios of these compounds \citep{prin93}.
Icy bodies more massive than $\sim 0.02$~$M_{\bigoplus}$ form with 
enough accretional energy that they begin life with their ice mostly
melted \citep{luni82, stev86}.  
These objects must freeze again in the outer solar system,
though the details of this process remain uncertain.

Jupiter's moon Ganymede has a mass of only 
0.02~$M_{\bigoplus}$, but it makes a good reference point \citep{sohl02}.
Ganymede has a diameter larger than that of Mercury (0.06~$M_{\bigoplus}$);
its mean density suggests that it is roughly 60\% silicate rock and metal
and 40\% volatiles. 
Water ice appears to cover most of Ganymede's surface; this
ice may top an ocean maintained by radiogenic heating \citep{spoh03}.
Objects even richer in water could possibly form near the snow line, where
water may collect due to diffusion in the
solar nebula \citep{stev88}.  

Interaction with a circumstellar disk can cause 
protoplanets to migrate inward from the snow line \citep{gold80,lin96}.
Ice planets could form with masses up to the threshold for rapid gas accretion
($\sim 10 M_{\bigoplus}$); low mass objects in this category 
undergo ``type I'' migration in a massive protoplanetary disk, which appears
to occur on timescales of $10^{5} (M/M_{\bigoplus})^{-1}$ years,
where $M$ is the mass of the planet \citep{ward97}.
If the volatile-rich planet migrates after the $\sim 10^7$~year period of
terrestrial planet formation at $\sim 1$~AU, it has a better
chance of surviving near the star.  Even after 
the outer protoplanetary disk has dissipated, interaction with other protoplanets
can scatter an icy body into an orbit with small pericenter \citep{rasi96,adam03},
where further encounters, stellar tides \citep{mayo95} or disk tides \citep{ward88} can
circularize its orbit.  Scattered Earth-mass planetesimals in
the outer solar system may have generated the obliquity of Uranus and Neptune \citep{ster91}
and the high inclination Kuiper Belt objects \citep{peti99}.

A volatile planet may not even need to migrate to get to the
habitable zone.
Recent models of flared disks
around young stars suggest that the snow line could be as close-in as
1~AU in a disk around a solar-mass star \citep{sass00}.
Asteroids orbiting between 2.6 and 3.5 AU (most C-type asteroids)
contain minerals that appear to have undergone aqueous alteration,
suggesting that they contained water ice when they formed
\citep[see the review by][]{baru95}.
A snowline located closer-in than 2.6 AU
would naturally explain the existence of
icy primitive asteroids.


\section{VAPORIZATION: A SELF-LIMITING PROCESS}

A water planet arriving at $\sim 1$~AU should quickly develop a 
steam atmosphere of 100 bars or more, like the atmosphere of the accreting Earth \citep{abe85}.
Such an atmosphere can easily become a runaway greenhouse when accretional impacts
deliver an energy flux $\gtrsim$ the solar constant to its base.
But in the absence of such an energy source, the two-stream model of 
\citet{mats86} suggests that a steam atmosphere of $\gtrsim 30$~bars
simply provides stable greenhouse heating up to a temperature of
\begin{equation}
T_{G}=\left({\lambda+\sqrt{1-\omega}} \over {1+\sqrt{1-\omega}} \right)^{1/4} T_{BB}
\approx \lambda^{1/4} T_{BB},
\end{equation}
at its base, where $\lambda$ is the ratio of the opacity at thermal wavelengths to the opacity
at short wavelengths, 
$\omega$ is the albedo for single-scattering at short wavelengths, and
$T_{BB}$ is the local blackbody temperature; 
$T_{BB}= 278 \ K \ (a/{\rm 1~AU})^{-1/2} (L_{*}/L_{\odot})^{1/4}$, where
$a$ is the radius of the planet's orbit, and $L_{*}$ is the luminosity of the star.
Since $\lambda$ is likely to be in the
range of 10--100  ($\lambda \approx 31$ for Venus), and $\omega$ is likely to be
almost 1, $T_{G} \approx 2.4 T_{BB}$.

When $T_{G} < 647$~K, 
the critical temperature of water, water simply condenses out of a steam atmosphere
that is in equilibrium with stellar radiation, leaving a hot ocean on the planet's surface
\citep{mats86, zahn88}.  
In this situation, the planet's vaporization halts, and the planet can survive
indefinitely as long as it retains its atmosphere.  
This condition holds where  
$a \gtrsim {\rm 1.03~AU} (L_{*}/L_{\odot})^{1/2}$, for $\lambda = 31$, $\omega=1$.

Closer in, the planet will slowly vaporize (or sublimate, if ice remains on its
surface) at a rate set by the rate of energy transport through the
base of the optically thick atmosphere, where the temperature will 
fall from $T_{G}$ to 647~K at the surface of the boiling ocean.
The atmosphere would be stable against
convection since it is heated from above; downward energy transport in this
stable layer will occur by slow diffusion of radiated thermal energy. 
The flux of energy
transported by radiative diffusion through the lowest scale
height of the atmosphere would be
$F_{diff} \approx -(16/3)\sigma T^3 \Delta T/\tau_R(T)$,
where $\sigma$ is the Stefan-Boltzmann constant and
$\tau_R(T)$ is the Rosseland mean optical depth of the layer.
The temperature difference across the layer, $\Delta T$
will be $\Delta T = T_{G} - 647$~K.  The mean temperature, $T$,
will be roughly $(T_{G}+647 \ \rm{K})/2$.

The more massive the atmosphere, the slower the vaporization; this process
is probably self-limiting.  The vaporization rate is
$\sim 4 \pi r_0^2  F_{diff}/L$, where $r_0$ is the radius of the planet, and
$L \approx 2.26 \times 10^{10}$ erg~g${}^{-1}$
is the latent heat of vaporization of water. 
The optical depth of the atmosphere will increase as more volatiles sublimate, but
when $\tau_{R} \approx 10^5 (a/1~AU)^{-2} (L_{*}/L_{\odot})$, the
vaporization rate reaches $\sim 10^{-10}$~${M}_{\bigoplus}$~yr${}^{-1}$,
i.e., it effectively halts.

Suspended aerosols in the atmosphere could easily provide a continuum
mass opacity of $\kappa = 0.1$~cm${}^2$~g${}^{-1}$,
like clouds in brown dwarfs \citep{coop03}.
The planet should have a radius similar to that of the
zero temperature sphere described by \citet{zapo69};
a zero temperature sphere with mean atomic mass per atom of 13~AMU
(half H${}_2$O, half SiO${}_2$)
has a radius of $r_0 \approx 1.4$~R${}_{\bigoplus} (M/{\rm M}_{\bigoplus})^{1/3}$.
With these assumptions, the limiting optical depth amounts to a mass of gas
\begin{equation}
M_{max}=2 \times 10^{-3} \ M_{\bigoplus} \ (\kappa / {\rm 0.1~cm}^2{\rm g}^{-1})^{-1} (a/1~AU)^{-2} (L_{*}/L_{\odot}) (M/M_{\bigoplus})^{2/3}.
\label{eq:maxmass}
\end{equation}

Though this thick atmosphere can protect the planet's volatiles from 
vaporization, the planet is vulnerable 
during the first billion years of the solar system's life,
when the star shines brightly in the EUV, and the planet suffers heavy
bombardment by planetesimals.  We will consider each of these processes in turn.

\section{EUV-DRIVEN ESCAPE}

Stellar EUV photons can raise the kinetic energy of gas
molecules above the escape velocity, eroding the planet.
For example, Pluto has probably lost a few km of surface ice during
its lifetime to an EUV driven wind \citep[see the review by][]{traf96}.
We can estimate the escape rate by estimating the radius,
$r_1$, where the atmosphere first becomes optically thick to EUV.
This parameter sets the effective cross-section of
the planet for absorbing stellar EUV and also the thickness
of the layer which must conduct the deposited energy to the surface.
We will neglect the vertical thickness of the opaque
region of the atmosphere just above the surface; its
scale height should be
$h_s \approx {59~\rm km} \ (M/ M_{\bigoplus})^{-2/3}$
(assuming steam at 647~K), much less than the
typical scale height of the upper layers.  This approximation 
works as long as most of the planet remains liquid (or ice) so the atmosphere
has only a small fraction of the planet's mass.

\citet{wats81} and \citet{hunt82} modeled the escape of Pluto's atmosphere
and the atmospheres of the primordial Earth and Venus
using a one-dimensional hydrodynamic model based on the
model for the solar wind by \citet{park63}.  This model estimates
the effective value of $r_1$ for the case of maximum escape
flux by assuming all the EUV energy is deposited in a thin layer at $r_1$.
This maximum escape flux provides
a lower limit for the lifetime of the bound gas.


\citet{wats81} define the variables
$\lambda_0 = GMm/{k T_0 r_0}$, $\beta = S (GMm/{kT_0^2 \kappa_0})$
and $\zeta = F_{esc} (k^2 T_0/{\kappa_0 GMm})$, where
$m$ is the mass of a gas molecule.
The base of the exosphere, which we will consider to be located
at $r_0$, has temperature $T_0$, which we take to be $2.4 T_{BB}$. 
The flux of escaping particles per steradian per second
is $F_{esc}$, and the effective EUV heating rate is
$S$, in erg~cm${}^{-2}$~s${}^{-1}$, averaged over a
sphere.  The thermal conductivity of the gas is
assumed to take the form  $\kappa=\kappa_0(T/T_0)^s$.
The maximum escaping flux satisfies \citep{wats81}:
\begin{equation}
\zeta = p \left({{(\lambda_1/2)^{{1/p} +1}} \over {\lambda_0 - \lambda_1}} \right)^2
\end{equation}
\begin{equation}
\lambda_1^2 = {\beta \over {\zeta \left (\lambda_0 - \left({p / \zeta} \right)^{1/2} \right)   }}  ,
\end{equation}
where $p = 2/(s+1)$.
We solved these equations numerically for $\zeta$ as a function of
$\beta$ and $\lambda_0$ and integrated the mass loss rate, $\dot M= 4 \pi m F_{esc}$,
over time, $t$, since the birth of the planetary system.
Since vaporization immediately replaces mass lost to the EUV-driven wind,
$\dot M$ is the mass evolution of the planet.

This hydrodynamic model applies when $\lambda_0 \gtrsim 10$,
i.e., the atmosphere is tightly bound.
As $\lambda_0$ decreases, 
EUV penetrates deeper into the atmosphere, and
$r_1$ approaches $r_0$ \citep[see figure 3 of][]{wats81}.
We assume that when $\lambda_0 < 7$, $r_1$=$r_0$, meaning
that the EUV penetrates all the way to where the atmosphere
becomes opaque near the surface of the planet.
In this case, $F_{esc} \approx S r_0^3/(GMm)$.

The thermosphere may have complicated photochemistry;
EUV can dissociate any of the key atmospheric molecules
(H${}_{2}$O, NH${}_3$, CH${}_4$, CO, N${}_2$,
and CO${}_2$).  But we can conservatively estimate the loss rate of
these compounds as the escape rate for monatomic H (m = $1.67 \times 10^{-24}$~g).
\citet{bank73} calculate that the thermal conductivity for H is
$\kappa=379 T^{0.69}$~erg~cm${}^{-1}$~s${}^{-1}$~K${}^{-1}$ between
$T=200$ and 2000~K.

The effective value of the present day EUV flux at the Earth
is $S \approx 1$~erg~cm${}^{-2}$~s${}^{-1}$ \citep{wats81}.
Younger solar-mass stars can easily have 100 times
as much EUV output as the sun.  We assume that 
$S = {\rm 1~erg~cm}^{-2}$~s${}^{-1} \ (a/{\rm 1~AU})$ 
at $t=4.5$~billion years and thereafter.
We assume that prior to that, the EUV output evolved as
$(t/4.5 \times 10^{9} \hbox{years})^{-0.8}$, up to
$S = {\rm 100~erg~cm}^{-2}$~s${}^{-1} \ (a / {\rm 1~AU})^{-2}$
following \citet{zahn82}.

\begin{figure}
\epsscale{0.8}
\plotone{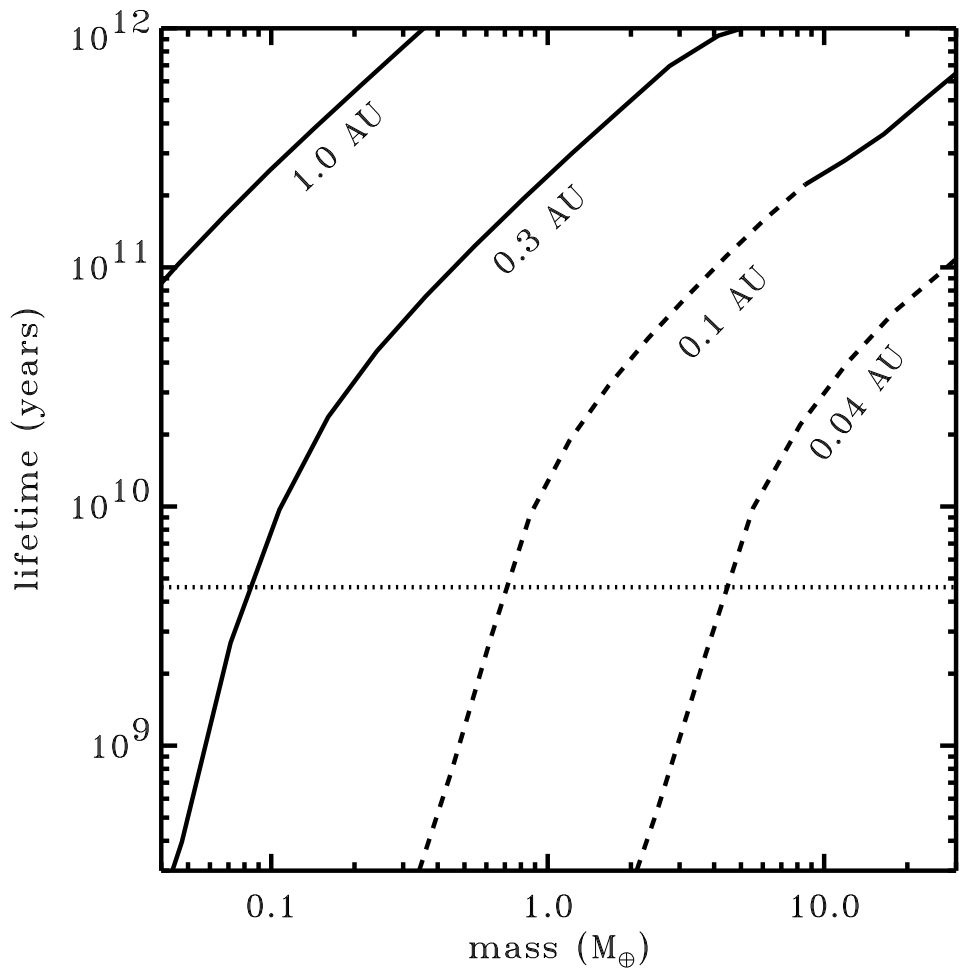}
\caption{Survival time for EUV-driven atmospheric escape
as a function of mass for volatile planets at
1~AU, 0.3~AU,  0.1~AU, and 0.04~AU.  The curves become dashed
where the atmosphere becomes $\gtrsim 10$\% of the planet's
mass. The dotted line shows the age of the solar system.}
\label{fig:survival}
\end{figure}

Figure~\ref{fig:survival} shows the EUV survival time---the time
for an all-volatile planet to lose half of its mass to an
EUV-driven wind---as a function of initial mass
for planets at $a=1$, 0.3, 0.1, and 0.04~AU.  
The figure suggests that an Earth-mass
volatile-rich planet as close as 0.3~AU from a sunlike star can
easily retain its volatiles for the age of the solar system.
Our model of EUV-driven escape
assumes that most of that planet,
say 90\%, remains ice or liquid, so that the base of the
thermosphere sits near the planet's surface.
We estimate that this situation occurs only where
$a \gtrsim {\rm 0.14~AU} (M/{M}_{\bigoplus})^{-1/6} (L_{*}/L_{\odot})^{1/2}$.
The dashed sections of the curves in Figure~\ref{fig:survival} indicate
the regime where this condition is not satisfied.


\clearpage

\clearpage

Since volatile-rich Earth-mass planets slowly lose mass to a EUV-driven wind,
these planets should be more common around younger
stars and stars with lower time-averaged EUV luminosity.
Nearby F-, G-, and K-type main-sequence stars have similar
levels of EUV output, but nearby M-type main-sequence stars have
systematically lower EUV luminosities than more massive nearby
main-sequence stars \citep{wood94}.  This difference
might make M-type stars particularly hospitable to close-in, volatile-rich
Earth-mass planets.


\section{BOMBARDMENT}

We assume that the volatile-rich planets migrate after the usual
period of terrestrial planet accretion in habitable zone.  
But ``heavy'' bombardment by leftover planetesimals
continued on the terrestrial planets in our solar system for roughly 
the first billion years of the solar system's life.
Such a bombardment can erode the atmosphere of a planet
\citep[see the review by][]{ahre93}.  

A low-mass volatile-rich planet will have escape velocity
$v_{esc} \approx {\rm 9~km~s}^{-1} \ ({\rm M}/{\rm M}_{\bigoplus})^{1/3}$.
Ejecting a mass $m_{e}$ of material from the planet 
requires energy ${1 \over 2} m_{e} v_{esc}^2$.
But the impactors
available in the early solar system would rarely have impact velocities more than
a few times $v_{esc}$, so an impactor probably could not eject more than
roughly its own mass from the planet's atmosphere,
given the $\sim$7\% efficient ejection process found in numerical
models of large impacts on the Earth \citep{newm99}.
Moreover, only impactors with diameter greater than the scale
height of the atmosphere impart a substantial fraction of their energy to the
atmosphere \citep{okee82}; smaller impactors just add mass to the planet.
A scale height of 59~km corresponds to a projectile with
mass $\sim 4 \times 10^{17}$~kg.

We appropriate a description provided by \citet{melo89} for the
flux of impactors on Mars and the Moon based
on crater densities.  The number of projectiles per year
of mass greater than $m_{p}$ that fall on a planet of effective radius $r_{eff}$
is
\begin{equation}
N_{cum}= 0.25 \ {\rm yr}^{-1} \ (r_{eff}/R_{\bigoplus})^2 \left (1 + 2300 e^{-t/2.2 \times 10^{8} \ {\rm yr}} \right)(m_{p}/1 {\rm kg})^{-0.47},
\label{eq:impactor}
\end{equation}
where we have neglected gravitational focusing---a factor of a few when scaling
from Mars to a 1~$M_{\bigoplus}$ planet.
If we consider only impactors with masses between $10^{15}$~kg and
1\% of the mass of the moon on a planet with $r_{eff}=R_{\bigoplus}$,
Equation~\ref{eq:impactor} implies a total mass of impactors of
less than 0.003~$M_{\bigoplus}$.
This total mass of impactors poses no hazard whatsoever, even
to the survival of an 0.1~$M_{\bigoplus}$ volatile-rich planet.

However, the above impactor distribution fails at the high mass end; it
does not account for the unusual impactors \citep{trem93} that have masses
comparable to a terrestrial planet.  Such giant impactors
probably created the Earth's moon \citep[e.g.,][]{ward78, canu01}
and may have spun up Mars \citep{done93} and robbed Mercury of its
low-density material \citep{benz88}.
Conceivably, a giant impactor with high enough velocity
could remove all the volatiles from a 1~$M_{\bigoplus}$ volatile planet.

\section{CONCLUSIONS}

The proximity of the snow line to 1 AU in accretion disk models 
suggests that volatile-rich planets can easily migrate into the
habitable zones of solar-type stars.
As long as they migrate after terrestrial planet formation and manage
to avoid giant impacts, we estimate that volatile-rich planets as small as
0.1~$M_{\bigoplus}$ can retain most of their volatiles for the lifetime
of the solar system as close as 0.3~AU from a solar-type star.
Closer-in objects may also retain volatiles, but our
analysis fails at high heating rates, where the atmosphere's mass becomes
substantial.
Exterior to $a \gtrsim {\rm 1.03~AU} \ (L_{*}/L_{\odot})^{1/2}$,
a water planet can exist in equilibrium with stellar radiation, protected by
a thick steam atmosphere.  Inside this radius, the planet slowly
vaporizes, but we estimate that only a mass 
of water given by Equation~\ref{eq:maxmass} can vaporize in $10^{10}$ years.
Conceivably, both the Earth and Mars could once have been volatile-rich,
but both have probably suffered large impacts.
Volatile-rich planets should be more common in the
habitable zones of young stars and M-type stars.

Volatile-rich Earth-mass planets could appear in future surveys for
extrasolar Earth analogs.  These objects should maintain deep layers of
liquid water, melted by accretional energy and radionucleides.  At a given mass,
a volatile-rich planet should have a slightly larger radius and therefore a
greater luminosity than a terrestrial planet.  
Finding such objects would test core-accretion scenarios for giant planet formation.
A small planet is not necessarily a terrestrial planet.

\acknowledgements

Thanks to Ann Bragg, Ulyana Dyudina, Mike Lecar, Dimitar Sasselov,
Sarah Stewart, and Wes Traub for discussions.
Thanks also to Alain Leger for sharing with me some of his unpublished
work on ocean planets.  This work was performed in part under contract
with the Jet Propulsion Laboratory (JPL)
through the Michelson Fellowship program funded by 
NASA as an element of the Planet Finder Program.  JPL
is managed for NASA by the California Institute of Technology.

\end{document}